\documentclass{article}

\usepackage{PRIMEarxiv}

\usepackage[utf8]{inputenc} % allow utf-8 input
\usepackage[T1]{fontenc}    % use 8-bit T1 fonts
\usepackage{hyperref}       % hyperlinks
\usepackage{url}            % simple URL typesetting
\usepackage{booktabs}       % professional-quality tables
\usepackage{amsfonts}       % blackboard math symbols
\usepackage{nicefrac}       % compact symbols for 1/2, etc.
\usepackage{microtype}      % microtypography
\usepackage{lipsum}
\usepackage{fancyhdr}       % header
\usepackage{graphicx}       % graphics
\graphicspath{{media/}}     % organize your images and other figures under media/ folder

\title{Antimatter Gravity and the Results of the ALPHA-g Experiment
}

\author{
  Massimo Villata \\
  INAF, Osservatorio Astrofisico di Torino, I-10025 Pino Torinese (TO), Italy \\
  \texttt{massimo.villata@inaf.it} \\
}

\begin{document}
\maketitle

\begin{abstract}

By combining general relativity and CPT symmetry, the theory of CPT gravity predicts gravitational repulsion between matter and CPT-transformed matter, i.e.\ antimatter inhabiting an inverted space-time. Such repulsive gravity turned out to be an excellent candidate for explaining the accelerated expansion of the Universe, without the need for dark energy. The recent results of the ALPHA-g experiment, which show gravitational attraction between antihydrogen atoms and the Earth, seem to undermine this success in the cosmological field. Analyzing the above theory, we find two solutions that can be consistent with the experimental results, while preserving the large-scale gravitational repulsion. The first highlights how repulsive gravity can be the result of the interaction with an inverted space-time, but occupied by matter and not antimatter, and therefore the antimatter present in our space-time has no reason to exhibit gravitational repulsion. The second retains the original CPT transformation, resulting in repulsive gravity between matter and antimatter, but with the caveat that antimatter immersed in our space-time cannot exhibit the PT transformation which is the cause of the repulsion. Finally, it is shown that, in a Newtonian approximation of the geodesic equation, time reversal is not a necessary operation for repulsive gravity, therefore opening the possibility of an expanding cosmos with a single time direction.
%&\lipsum[1]

\end{abstract}

% keywords can be removed
\keywords{Universe evolution \and Gravitational repulsion and antiparticles \and Baryon symmetry of the Universe}

\section{Introduction}

In 1998, astronomers discovered the acceleration of the Universe expansion (e.g.\ \cite{Rie1998,Per1999}). Since then, one of the most debated problems in physics and cosmology has been the existence and nature of the so-called dark energy, which should explain that unexpected finding. In fact, a repulsive force that acts in the Universe space-time challenges any previous physical knowledge, since the only known interaction among matter on these large scales is the universal gravitational attraction.

The cosmic expansion acceleration, i.e.\ the presence of dark energy, has been formally attributed to an additional term that introduces negative pressure into the equations governing cosmic expansion. In its simplest form, this term corresponds to a cosmological constant, which can be associated with the energy of the quantum vacuum. In addition to this standard cosmological model, known as the $\Lambda$CDM model, various alternative explanations have been proposed to account for the accelerated expansion. These alternatives involve resort to scalar fields or the application of modifications to general relativity, including extensions to extra dimensions or higher-order curvature terms (e.g.\ \cite{Ame2000,Dva2000,Car2004,Cap2005,Nap2012}).

Both the classical (Newtonian) and relativistic (Einsteinian) gravitation theories seem to exclude that gravity could be in any way repulsive. However, it has been shown that general relativity can be consistently extended to the presence of antimatter (which was unknown at the time of the formulation of the two theories), based on its CPT properties, which imply that matter and antimatter are both gravitationally self-attractive, but mutually repulsive \cite{vil2011}. Therefore, if our Universe contains a certain amount of antimatter (perhaps equivalent to that of matter, due to the expected matter-antimatter symmetry, and possibly located in cosmic voids), the origin of the cosmic acceleration can be explained easily and naturally (together with the well-known expansion itself), with no need for unknown ingredients such as dark energy, or changes to current consolidated theories \cite{vil2013}.

A particular case where general relativity can further support repulsive gravity between matter and antimatter is the Kerr space-time. In fact, in \cite{vil2015}, it was shown that the CPT transformation requested in ${\bf R}^4$ to get matter-antimatter gravitational repulsion is equivalent to the presence of the negative-$r$ region in the ${\bf R}^2\times S^2$ topology of Kerr(-Newman) space-time, which can be interpreted as the habitat of antimatter.

Cosmological models built on this theory of repulsive gravity between matter and CPT-transformed matter (in short CPT gravity) have shown many advantages with respect to current dark-energy models. They are based only on well-known physical entities and theories, with no need for ad-hoc, unknown but dominant, components. Moreover, they can spontaneously solve several heavy issues like the horizon and coincidence problems, the initial singularity, the apparent matter-antimatter asymmetry \cite{vil2013}.

In addition, several observed properties of our extragalactic neighborhood, such as the extreme emptiness of the Local Void, the unexpected presence of large galaxies at its edge and other anomalies, indicate the existence of a mechanism that can produce a fast evacuation of voids and rapid assembly of matter into structures, which the standard theory based on attractive-only gravity cannot provide. Repulsive gravity can be the cause of fast evacuation and, also through the squashing tidal effect, can help a lot to rapidly gather matter into galaxies and structures, especially at the borders of voids \cite{vil2012b}. In that paper, the author, upon conducting straightforward dynamical analyses, determined that the Local Void has the potential to accommodate a substantial quantity of antimatter, approximately equal to $5\times10^{15}\,M_\odot$. This quantity is roughly equivalent to the mass of a typical supercluster, thereby restoring a balance between matter and antimatter within the Universe. The repulsive gravity field generated by this entity, referred to as the ``dark repulsor'', offers a plausible explanation for the peculiar motion of the Local Sheet, which moves away from the Local Void.

On a larger cosmological scale, the gravitational repulsion originating from the presence of antimatter concealed within cosmic voids can supply more than adequate potential energy to propel both the ongoing Universe expansion and its acceleration. Importantly, this negates the necessity for an initial ``explosion'' or the introduction of dark energy as an explanatory factor. Additionally, the discrete distribution of these dark repulsors, in contrast to the uniformly pervasive dark energy, serves as a plausible explanation for dark flows and the observations of excessive inhomogeneities and anisotropies in the structure of the Universe.

Other researchers have worked on cosmological models based on matter-antimatter gravitational repulsion (e.g.\ \cite{Haj2011,Haj2014,Haj2020,Ben2012,Ch2018}), which was obtained by assuming that the gravitational mass of antimatter is (at least partially) negative, and not, as in CPT gravity, with masses that are always strictly positive.

In a recent paper (\cite{Dim2023}), N-body simulations were performed in the context of a matter-antimatter Universe with repulsive gravity between the two components. The Hubble expansion law is obtained, together with the acceleration of the Universe expansion, with no need for a dark-energy component.

However, all this success of the CPT-gravity theory (and similar matter-antimatter repulsive-gravity theories) is seriously undermined by the recent results of the ALPHA-g experiment, which show gravitational attraction of antihydrogen atoms in the Earth's gravity field \cite{alpha2023}. This can rule out the assumption that the gravitational mass of antimatter is negative. What we want to investigate here is what is wrong with CPT gravity (where the mass is always strictly positive), and find a solution that brings theory into agreement with experiments.

%%%%%%%%%%%%%%%%%%%%%%%%%%%%%%%%%%%%%%%%%%
\section{Analysis of the theory of CPT gravity}
\label{sec:analysis}

We consider here the theory of CPT gravity as developed in \cite{vil2011}, also recalling some useful concepts.

The CPT transformation is a fundamental concept in theoretical physics, particularly in the field of quantum field theory and particle physics. It refers to a combination of three discrete operations:

{\it Charge conjugation} (C): This operation involves changing all particles into their corresponding antiparticles while keeping their positions and momenta in space and time unchanged. In other words, it transforms particles with a certain electric charge into their antiparticles with opposite electric charge.

{\it Parity transformation} (P): Parity refers to the spatial inversion or mirror reflection of a physical system. The parity transformation flips the spatial coordinates, effectively changing left-handed particles into right-handed particles (or vice versa) without altering other properties.

{\it Time reversal} (T): This operation involves reversing the direction of time, essentially changing the direction of particle trajectories from past to future or vice versa.

The CPT transformation combines these three operations. According to the CPT theorem \cite{Lud1957}, the combination of charge conjugation, parity transformation, and time reversal leaves the fundamental laws of physics invariant for all known particles and their interactions. This implies that if you observe a physical process and then apply the CPT transformation to it, you should see the same process occur in reverse, with all particles transformed into their corresponding antiparticles, positions and momenta inverted in space, and time running backward.

The CPT symmetry is a crucial principle in modern physics, and any violation of this symmetry would have profound implications for our understanding of the fundamental laws of the Universe. So far, experimental observations have shown a high degree of agreement with the CPT symmetry, reinforcing its significance in particle physics.

For classical particles and antiparticles, the CPT transformation is
\begin{equation}
\label{eq.1}
{\rm CPT\,:}\qquad{\rm d}x^\alpha\,\rightarrow\,-{\rm d}x^\alpha\,,\qquad q\,\rightarrow\,-q\,,%\eqno(1)
\end{equation}
where $q$ is the electric charge.

The PT part of transformation (\ref{eq.1}), i.e.\ ${\rm d}x^\alpha\rightarrow -{\rm d}x^\alpha$, changes the sign of each component of any four-vector and odd-rank tensor, while it will be ineffective on even-rank tensors, being applied an even number of times. When the whole CPT transformation is applied, tensors containing the electric charge will suffer an additional sign change, so that even-rank (odd-rank) tensors, which are PT-even (PT-odd), become CPT-odd (CPT-even) in this case.

The gravitational equation of motion of general relativity, i.e.\ the geodesic equation, is
\begin{equation}
\label{eq.2}
{{\rm d}^2x^\alpha\over{\rm d}\tau^2}=-{\Gamma^\alpha}_{\beta\gamma}{{\rm d}x^\beta\over{\rm d}\tau}{{\rm d}x^\gamma\over{\rm d}\tau}\,,
\end{equation}
where the affine connection ${\Gamma^\alpha}_{\beta\gamma}$ represents the gravitational field and the other three elements refer to the test particle. Each element is CPT-odd. If we CPT-transform all the four elements, we obtain an identical equation describing the motion of an antimatter test particle in an antimatter-generated gravitational field, since all the four changes of sign cancel one another. Thus, this CPT symmetry ensures the same self-attractive gravitational behavior for both matter and antimatter. However, if we transform only one of the two components, either the field ${\Gamma^\alpha}_{\beta\gamma}$ or the particle (represented by the remaining three elements), we get a change of sign that converts the original gravitational attraction into repulsion, so that matter and antimatter repel each other.

As a consequence of the equivalence principle, in the equation of motion the mass disappears. However, in the following equation it can be useful to keep the ratio $m_{({\rm g})}/m_{({\rm i})}=1$ visible. For a useful comparison with electrodynamics, we also add the Lorentz-force law:
\begin{equation}
\label{eq.3}
{{\rm d}^2x^\alpha\over{\rm d}\tau^2}=-{m_{({\rm g})}\over m_{({\rm i})}}{{\rm d}x^\beta\over{\rm d}\tau}{\Gamma^\alpha}_{\beta\gamma}{{\rm d}x^\gamma\over{\rm d}\tau}+{q\over m_{({\rm i})}}{F^\alpha}_\beta{{\rm d}x^\beta\over{\rm d}\tau}\,.
\end{equation}
Here too all the elements are CPT-odd, since the first two elements of the Lorentz-force law contain an electric charge, and therefore this force reverses when one of the two components (particle or field) is CPT-transformed, as is commonly expected from the C transformation alone.

By comparing the gravitational part with the electrodynamic one, it can be seen how the electric charge $q$ of the latter corresponds to a gravitational charge $m_{({\rm g})}{\rm d}x^\beta/{\rm d}\tau$, i.e.\ the energy-momentum four-vector $p^\beta$, which is CPT-odd. This highlights how the common simplistic opinion that the gravitational charge is the mass (which is always positive and therefore gravity would always be attractive) is incorrect, while $p^\beta$ changes sign in an inverted space-time, thus giving repulsive gravity.

%%%%%%%%%%%%%%%%%%%%%%%%%%%%%%%%%%%%%%%%%%
\section{Possible solutions to the conflict with ALPHA-g results} 

We note that in Equations (\ref{eq.2}) and (\ref{eq.3}) repulsive gravity is obtained by the PT transformation of one of the two components (particle or field), while C is ineffective. This could mean that the repulsion occurs between two opposite space-times, but nothing guarantees that one of the two is necessarily populated with antimatter. The presence of antimatter is required only by CPT symmetry, where the three transformations must occur together.

This solution without antimatter would save the cosmological scenario where alternate ``islands'' of opposite space-times cause the needed gravitational repulsion to account for the accelerated expansion of the Universe. But it would break the matter-antimatter symmetry of the cosmological model in \cite{vil2013}.

On the other hand, the Lorentz-force law in Equation (\ref{eq.3}) is sensitive only to C (changing the sign of $q$ and/or ${F^\alpha}_\beta$), while PT is ineffective, since the two possible changes of sign of the four-acceleration and four-velocity would cancel each other out. So nothing guarantees that antimatter is in an inverted space-time. Apart from the requirement of CPT symmetry.

In fact, in experiments with antimatter in our laboratories, nothing seems to reveal an inverted space-time for it. In particular, antimatter appears to follow our time arrow well, and the ALPHA-g results of non-repulsive gravity seem to confirm the lack of PT inversion.

All this could therefore be a solution to the conflict between CPT gravity and CERN results. It preserves the gravitational repulsion between opposite space-times needed for accelerated cosmic expansion, but without requiring matter-antimatter repulsive gravity. Although the price to pay is the lack of matter-antimatter symmetry in the cosmological model.

Another solution, which preserves both repulsive gravity and cosmological matter-antimatter symmetry, is the following.

CPT gravity is the original one (C and PT operate together), and therefore an inverted space-time contains antimatter.

But what about the tiny amounts of antimatter in our laboratories or in our space-time surroundings? As already mentioned, it seems that they belong to our space-time. And couldn't this mean that two opposing space-times cannot coexist? But that the one that is by far dominant can cancel out the other? Just think of the causality, thermodynamic and entropic conflict that would arise with two overlapping arrows of time (for similar problems see e.g.\ \cite{Sch1999} and references therein). And the evident symptom of this could not be the immediate annihilation of antimatter, momentarily forced to follow an arrow of time that is not its own?

If the answer is yes, then the antimatter we know, which travels forward in time and does not exhibit gravitational repulsion, is not like the CPT-transformed one which occupies islands of space alternating with those of matter and causes accelerated expansion, as shown in \cite{vil2013}. And gravitational repulsion would be just the mechanism necessary to ensure that islands and anti-islands never meet, and that the coexistence of two opposite space-times does not cause a cosmic annihilation.

\begin{figure}%&&[H]
%&&\begin{adjustwidth}%&&{-\extralength}%&&{0cm}
\centering
\includegraphics[width=15.5cm]{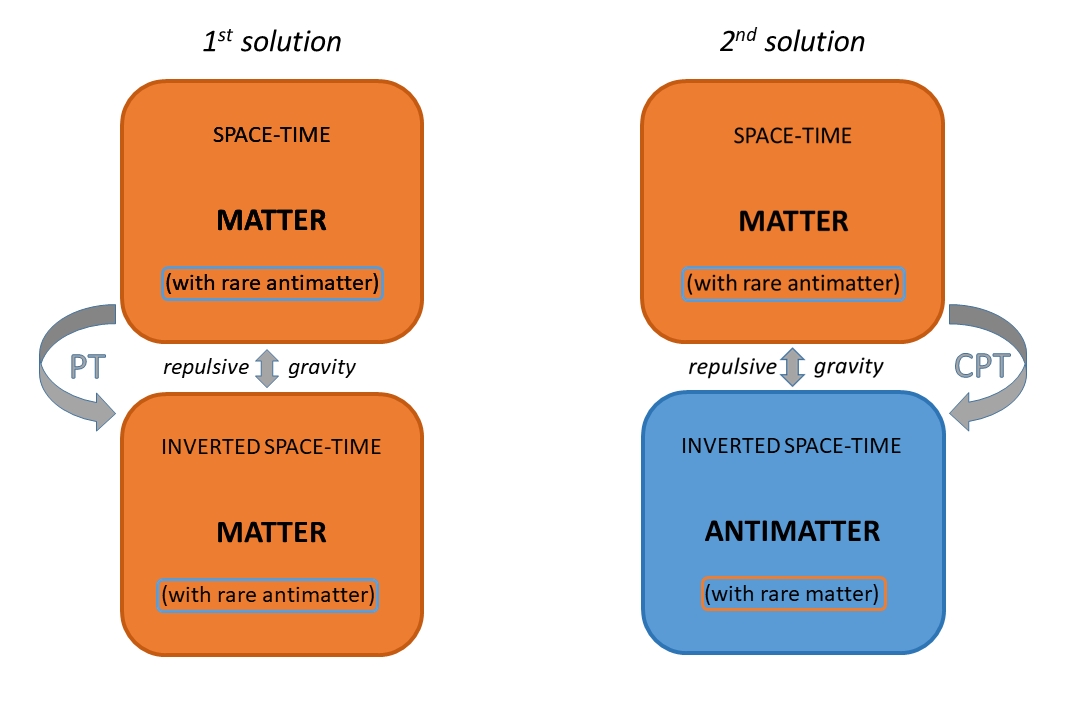}
%&&\end{adjustwidth}
\caption{Schematization of the two solutions to the conflict between CPT gravity and ALPHA-g results. In the first case (on the left), the large-scale repulsive gravity is given by the interaction with PT-transformed matter and not with antimatter, in a matter-dominated Universe. In the second case (on the right), the entire CPT is conserved, giving matter-antimatter repulsive gravity. But the tiny amount of antimatter immersed in our space-time cannot be PT-transformed (see text for further explanation).\label{fig1}}
\end{figure}

The two solutions described above are schematized in Figure \ref{fig1}.
% Example of a figure that spans the whole page width. The same concept works for tables, too.

%%%%%%%%%%%%%%%%%%%%%%%%%%%%%%%%%%%%%%%%%%
\section{Discussion and conclusions} 

Unfortunately, astronomical observations cannot distinguish between the two solutions, since in both cases the radiation emitted from an inverted space-time would be received by us as advanced radiation, for which we do not (yet) have adequate detectors. And this is the reason why cosmic voids are the best candidates to host (C)PT-transformed matter, both for the absence of radiation and for their evident repulsive gravity in the confinement of clusters and superclusters of galaxies, and to explain the properties and anomalies of our extragalactic neighborhood, due to the proximity of the Local Void.

We note that the ALPHA-g experiment yields a value of (0.75 ± 0.13 (statistical + systematic) ± 0.16 (simulation)) $g$ for the local acceleration of antimatter towards the Earth, which does not guarantee a value exactly equal to $g$, as one would expect. If more precise future results show an acceleration significantly less than $g$, one might wonder whether this is due to a partial presence of the inverted space-time (of the second solution above) that was not completely destroyed.

There are many people who distrust repulsive gravity because it would not be consistent with the equivalence principle. The fact is that the equivalence principle, in its various forms, is implicitly considered valid when the masses and fields involved belong to the same space-time. Repulsive gravity arises from general relativity when masses and fields belong to two opposite space-times, and therefore it too is a consequence of the equivalence principle. However, this should perhaps be reformulated to avoid such misunderstandings.

Even the representation of space as a sheet deformed by the presence of masses, which therefore deviates the test particle from its rectilinear trajectory, is easy to understand for repulsive gravity. The test particle in an opposite space-time is ``below'' the sheet, and perceives the deformations produced by the masses as repulsive hills.

In summary, there are two solutions, consisting of amendments to the original CPT-gravity theory, which allow to conserve the large-scale repulsive gravity needed for the accelerated expansion of the Universe and at the same time to get along with the results of the ALPHA-g experiment. The first solution must give up the matter-antimatter symmetry of the original theory, separating the CPT transformation into independent C and PT ones. That is, antimatter does not belong to an opposite space-time and in the opposite space-time antimatter does not dominate. The second preserves the cosmic matter-antimatter symmetry and the entire CPT symmetry, with the caveat that the tiny quantities of antimatter immersed in our space-time must submit to it in their brief existence.

However, when faced with these two solutions, one might feel embarrassed by the presence of time reversal T, which is difficult to digest, especially in a cosmic context (but see e.g.\ \cite{Sch2005} and references therein). Then let's see if it is really necessary.

For this purpose, let's consider for simplicity the low-velocity, stationary-field approximation of the geodesic equation:
\begin{equation}
\label{eq.4}
{{\rm d}^2x^\alpha\over{\rm d}\tau^2}=-\Gamma^{\alpha}_{00}{{\rm d}x^0\over{\rm d}\tau}{{\rm d}x^0\over{\rm d}\tau}={1\over 2}g^{\alpha\beta}{\partial g_{00}\over\partial x^\beta}\left({{\rm d}t\over{\rm d}\tau}\right)^{\!2}\,,%\eqno(10)
\end{equation}
where part of the time inversion vanishes in the square of ${\rm d}t/{\rm d}\tau$.

If we then add the further assumption that the field is weak, we can easily find the vectorial form of the Newton law ${\rm d}^2{\rm\bf x}/{\rm d}t^2=-\nabla\phi$, which is completely insensitive to T, while the P-oddness of the field is in the gradient, and the P-oddness of the particle is in the acceleration.

So we can conclude that, at least in the Newtonian approximation, the equation of motion retains the same attractive/repulsive properties as the original geodesic equation, but now repulsive gravity could act between two opposite spaces, without time reversal.

As a result, we can place alongside the two solutions described above (PT and CPT gravity), two other solutions that do not require time inversion, thus opening the possibility that the cosmic repulsion is due to alternating islands of opposite spaces (with or without antimatter) with the same time direction.

%&\section*{Acknowledgments}
%&This was was supported in part by......

%Bibliography
\bibliographystyle{unsrt}  
\bibliography{references}

\end{document}